\documentclass[showpacs, floatfix,twocolumn]{revtex4}
\usepackage{amsmath}
\usepackage{color}
\usepackage{amssymb}

\usepackage{graphicx}

\begin{document}

\title{ Superconductivity in the Ferromagnet URhGe under uniaxial pressure}

\author{V.P.Mineev}
\affiliation{Univ. Grenoble Alpes, CEA, INAC, PHELIQS, GT, F-38000 Grenoble}

\begin{abstract}
Uniaxial pressure  applied  in the b crystallographic direction perpendicular to spontaneous magnetization
 in heavy fermion ferromagnet URhGe strongly stimulates superconductivity in this compound. 
The phenomenological approach allows point out two mechanisms of superconducting temperature raising. They originates from stimulation by the uniaxial stress  both intraband and interband amplitudes  of triplet Cooper pairing. The phenomenon of reentrant superconductivity under magnetic field along b-axis is also strongly sensitive to the uniaxial stress in the same direction. The uniaxial stress accelerates  suppression
the Curie temperature by the transversal magnetic field. The emergence of the first order transition to the paramagnetic state occurs at much lower field  than in the absence of uniaxial stress.

\end{abstract}

\date{\today}
\maketitle
\section{Introduction}
The coexistence of superconductivity and ferromagnetism  is the hallmark of heavy fermion uranium compounds UGe$_2$, URhGe and UCoGe  (see  the recent experimental \cite{Aoki14} and theoretical \cite{Mineev2016} reviews and references therein). The emergence of the superconducting state at temperatures far below the Curie temperature and very high upper critical field  strongly indicate on the spin-triplet Cooper pairing in these materials. One of the most peculiar observations is the phenomenon of reentrant superconductivity  in URhGe \cite{Levy2005} which is an orthorhombic ferromagnet with spontaneous magnetization along  c-axis. At low enough temperature  the magnetic field about 1.3 Tesla 
 directed along the $b$-axis suppresses the superconducting state \cite{Hardy2005} but at much higher field  about 10 Tesla the superconductivity is recreated and exists 
 till the field about 13 Tesla \cite{Levy2005}. The  maximum of the superconducting critical temperature in this field interval is $\approx 0.4~K$. In the same field interval the material transfers from the ferromagnet to the paramagnet state by means of the first-order type transition. The superconducting state exists not only inside of the ferromagnetic state but also in the paramagnetic state separated from the ferromagnetic state by the phase transition of the first order. The theoretical treatment of this phenomenon has been proposed in Ref.5. 
 
 There was shown experimentally that a hydrostatic pressure applied to URhGe crystals stimulates ferromagnetism causing an increase of the  Curie temperature $T_c( P )$  and,  at the same time, suppresses the superconducting state  decreasing the critical temperature of superconducting phase transition $T_{sc}( P )$ \cite{Hardy},
 as well as the maximum of the superconducting critical temperature of the reentrant superconducting state \cite{Miyake2009}.  The latter is also shifted to a bit higher field interval.
Quite  the opposite behavior has been registered recently \cite{Braithwaite2016} under uniaxial stress $P_y$ in b-direction. In what follows we shall use the $x,y,z$ coordinate axes  pinned correspondingly to the $a,b,c$ crystallographic directions. 
Namely, the uniaxial stress suppresses the ferromagnetism decreasing the Curie temperature $T_c( P_y)$ and stimulate the superconducting state  such that the temperature of superconducting transition increases so strongly 
that leads to the coalescence
 the superconducting and reentrant superconducting regions in $(H_y,T)$ phase diagram
already at quite moderate  uniaxial stress values.  The comparison  of $(H_y,T)$ phase diagrams at ambient pressure  and at some uniaxial stress is presented in Fig.1.

Here, I  show  that the stimulation of superconducting state originates from two mechanisms:  the Curie temperature suppression
stimulating the intraband pairing and the increase of the magnetic susceptibility along b-direction
stimulating interband pairing.
Then, making use the phenomenological approach developed in Re.5, I  consider the $(H_y,T)$ phase diagram modification caused by uniaxial stress along b-direction.  It is demonstrated  that the uniaxial stress strongly accelerates the process of  Curie temperature suppression by magnetic field along b direction. Also  the field induced transformation of the second to the first order
 ferro-para phase transition occurs  at much lower field $H_y^{cr}( P_y)$ values than in a stress absence. This leads to the coalescence of the superconducting and reentrant superconducting states.

\section{Free energy}
The Landau free energy density of an orthorhombic ferromagnet 
in
magnetic field ${\bf H}({\bf r})$  under an external pressure consists of magnetic, elastic and magneto-elastic parts
\begin{equation}
F=F_M+F_{el}+F_{Mel},
\end{equation}
where in the  magnetic part \cite{Mineev2016}
\begin{eqnarray}
F_M=\alpha_{z}M_{z}^{2}+\beta_{z}M_{z}^{4}+\delta_zM_z^6+\alpha_{y}M_{y}^{2}+\alpha_{x}M_{x}^{2}~~\nonumber\\
 +\beta_{xy}M_x^2M_y^2+\beta_{yz}M_{z}^{2}M_{y}^{2}+\beta_{xz}M_{z}^{2}M_{x}^{2}-{\bf M}{\bf  H},
\label{F11}
\end{eqnarray}
we bear in mind the orthorhombic anisotropy and also the term of the sixth order in powers of $M_z$. Here,  $x, y, z$ are the coordinates
 pinned to the $a, b, c$
crystallographic directions correspondingly, 
\begin{equation}
\alpha_{z}=\alpha_{z0}(T-T_{c0}),~~~~ \alpha_x>0,~~~~ \alpha_y>0. 
\end{equation}
The elastic part of free energy in an orthorhombic crystal is \cite{TheorEl}
\begin{eqnarray}
F_{el}=\frac{1}{2}\lambda_xu_{xx}^2+\frac{1}{2}\lambda_yu_{yy}^2+\frac{1}{2}\lambda_zu_{zz}^2\nonumber\\
+\lambda_{xy}u_{xx}u_{yy}+\lambda_{xz}u_{xx}u_{zz}+\lambda_{yz}u_{yy}u_{zz}\nonumber\\
+\frac{1}{2}\mu_{xy}u_{xy}^2+\frac{1}{2}\mu_{xz}u_{xz}^2+\frac{1}{2}\mu_{yz}u_{yz}^2,
\label{el}
\end{eqnarray}
where $u_{xx},u_{yy},u_{zz},u_{xy},u_{xz},u_{yz}$  are the components of deformation  tensor.

In constant magnetic field ${\bf H}=H_y\hat y$ the $x$-projection of magnetization $M_x=0$, hence, one can take into account just $y$ and $z$ magneto-elastic terms
\begin{eqnarray}
F_{Mel}=(p_xu_{xx}+p_yu_{yy}+p_zu_{zz})M_z^2\nonumber\\
+(q_xu_{xx}+q_yu_{yy}+q_zu_{zz})M_y^2 +ru_{yz}M_yM_z.
\label{Mel}
\end{eqnarray}

To find the deformaition arising under infuence of uniaxial stress  $P_y$ applied along $b$-axis one must solve the linear equations 
\begin{eqnarray}
\frac{\partial (F_{el}+F_{Mel})}{\partial u_{xx}}=0,~~~\frac{\partial (F_{el}+F_{Mel})}{\partial u_{zz}}=0,\nonumber\\
\frac{\partial (F_{el}+F_{Mel})}{\partial u_{yz}}=0~~~\frac{\partial (F_{el}+F_{Mel})}{\partial u_{yy}}=-P_y
\label{Py}
\end{eqnarray}
in respect of components of deformation tensor. 
Whereas  in the case of  hydrostatic pressure $P$ the corresponding system of equations is
\begin{eqnarray}
\frac{\partial (F_{el}+F_{Mel})}{\partial u_{xx}}=-P,~~~\frac{\partial (F_{el}+F_{Mel})}{\partial u_{zz}}=-P,\nonumber\\
\frac{\partial (F_{el}+F_{Mel})}{\partial u_{yz}}=0~~~\frac{\partial (F_{el}+F_{Mel})}{\partial u_{yy}}=-P
\label{P}
\end{eqnarray}

Solving the equations (\ref{Py}) and substituting the solution back to the sum 
{$F_{el}+F_{Mel}$ given by equations (\ref{el}) and  (\ref{Mel}) we obtain
the magnetic field $H_y$ and the uniaxial pressure $P_y$ dependent magnetic part of free energy density 
 \begin{eqnarray}
F_M=(\alpha_{z}+A_zP_y)M_{z}^{2}+\beta_{z}M_{z}^{4}+\delta_zM_z^6\nonumber\\
 +(\alpha_{y}+A_yP_y)M_{y}^{2}
 +\beta_{yz}M_{z}^{2}M_{y}^{2}
 -M_yH_y.
\label{F22}
\end{eqnarray}
The coefficients $\beta_z$ and $\beta_{yz}$ in this expression differ from corresponding coefficients
in Eq. (\ref{F11}). However, this difference is pressure independent
so long the magneto-elastic coupling has the simple form given by eq.(\ref{Mel}). Hence, in what follows, we keep for these coefficients the same notations as in Eq.(\ref{F11}).

The coefficient
at $M_z^2$ now is
 \begin{equation}
 \alpha_{z}(P)=\alpha_{z0}(T-T_{c0})+A_zP_y,
 \end{equation}
and the Curie temperature    
acquires the pressure dependence
\begin{equation}
T_{c0}(P_y)=T_{c0}-\frac{A_zP_y}{\alpha_{z0}}.
\label{curie}
\end{equation}
The expression of coefficients $A_z$ and $A_y$  through elastic and magneto-elastic moduli  (see Appendix)  don't give up hopes to get  a theoretical estimation of their value. But in fact  this is not 
necessary.
 As we shall see the important role play the sign of them  which is determined experimentally.
 The Curie temperature decreases with uniaxial pressure  \cite{Braithwaite2016} what
corresponds to positive  
  $A_z$ coefficient.

The coefficient
at $M_y^2$  also acquires the pressure dependence
 \begin{equation}
 \alpha_{y}( P_y)=\alpha_{y}+A_yP_y.
 \end{equation}
 The equilibrium magnetization projection along the $y$ direction is obtained by minimization of free energy (\ref{F22}) in respect of $M_y$
\begin{equation}
M_y=\frac{H_y}{2[\alpha_y(P_y)+\beta_{yz}M_z^2]}.
\label{My}
\end{equation}
 The measurements \cite {Braithwaite2016} 
shows that the susceptibility along $y$ direction $\chi_{y} (P_y)=\frac{M_y}{H_y}$ increases with pressure enhancement.
This is owing to both $T_{c0}(P_y)$ and much stronger  $\alpha( P_y )$ decrease with uniaxial pressure. So,  the coefficient $A_y$ is proved to be negative, thus
\begin{equation}
 \alpha_{y}( P_y)=\alpha_{y}-|A_y|P_y.
 \label{susc}
\end{equation}

We shall see in the next Section that 
both dependencies (\ref{curie}) and (\ref{susc}) cause the enhancement of superconducting critical temperature.  Then I will demonstrate that b-direction magnetic softening 
makes  easier the 
suppression of ferromagnetic state by magnetic field in $y$ direction and stimulates  
 the emergency of reentrant superconducting state.

The  hydrostatic pressure  creates the coefficients dependencies  of the same form
\begin{eqnarray}
\alpha_{z}(P)=\alpha_{z0}(T-T_{c0})+A^h_zP,\\
 \alpha_{y}(P)=\alpha_{y}+A^h_yP.
\end{eqnarray}
This case, however,
\begin{equation}
A^h_z<0,~~~A^h_y>0
\end{equation}
what corresponds, as we shall see,  to    stimulation of  ferromagnetism  and to suppression 
of superconductivity reported in \cite{Hardy,Miyake2009}. 

\section{Superconducting critical temperature}

The superconducting state  in two band (spin-up and spin-down) orthorhombic  ferromagnet is described  in its simplest form in terms of 
two complex order parameter amplitudes  \cite{Mineev2011},\cite{Mineev2016}
\begin{eqnarray}
&\Delta^\uparrow({\bf k},{\bf r})=\hat k_x\eta_\uparrow({\bf r})
,\nonumber\\
&\Delta^\downarrow({\bf k},{\bf r})=\hat k_x\eta_\downarrow({\bf r})
\label{A'}
\end{eqnarray}
 depending on the Cooper pair centre of gravity coordinate ${\bf r}$ and the momentum ${\bf k }$ of pairing electrons.  This particular order parameter structure allows to explain the specific temperature dependence of the upper critical field anisotropy in URhGe \cite{Hardy2005}, \cite{Mineev2016}.

 The corresponding critical temperature of transition to the superconducting state is determined by the BCS-type formula
\begin{equation}
T=\varepsilon~exp\left (-\frac{1}{g}  \right ),
\label{21}
\end{equation}
where the constant of interaction
\begin{equation}
g=\frac{g_{1x}^\uparrow+g_{1x}^\downarrow}{2}+\sqrt{\frac{(g_{1x}^\uparrow-g_{1x}^\downarrow)^2}{4}+g_{2x}^{\uparrow}g_{2x}^{\downarrow}}
\label{g}
\end{equation}
is expressed through the constants of intra-band pairing $g_{1x}^\uparrow,~g_{1x}^\downarrow$ and the constants of inter band pairing $g_{2x}^\uparrow,~g_{2x}^\downarrow$.
They are  functions of temperature, pressure and  magnetic field. Thereby the formula (\ref{21}) is, in fact, an equation for the determination of the critical temperature of the transition to the superconducting state.

The constants of intra-band pairing interaction for spin-up and spin-down bands 
are proportional to the averaged over the Fermi surface density of states  and the odd-part of susceptibility along the direction of spontaneous magnetization
( see  \cite{Mineev2016} Eq.(103))
\begin{eqnarray}
g_{1x}^\uparrow\propto\frac{
\langle \hat k_x^2N_0^\uparrow\rangle}{(2\beta_zM_z^2+\gamma^zk_F^2)^2},\\
g_{1x}^\downarrow\propto\frac{
\langle \hat k_x^2N_0^\downarrow\rangle}{(2\beta_zM_z^2+\gamma^zk_F^2)^2}.
\end{eqnarray}
 The magnetization below the Curie temperature is given by
 \begin{equation}
 2\beta_zM_z^2=\alpha_{z0}(T_{c0}(P_y)-T).
 \end{equation}
Assuming that at temperatures far below from $T_{c0}(P_y)$ this formula is still qualitatively valid we obtain using Eq.(\ref{curie})
\begin{equation}
 2\beta_zM_z^2\approx\alpha_{z0}(T_{c0}-A_zP_y).
 \label{Mz}
 \end{equation}
 Thus, the magnetization decreases with uniaxial pressure, what causes in its turn the increase of superconducting interaction constant.
 
 On the other hand, the constants of inter-band pairing interaction are determined by the difference of the odd parts  susceptibilities in $x$ and $y$ directions 
(see  \cite{Mineev2016} Eq.(104))
   \begin{widetext}
 \begin{eqnarray}
g_{2x}^\uparrow\propto\langle \hat k_x^2N_0^\uparrow\rangle
\left\{\frac{\gamma_{xx}^x}{(\alpha_x+\beta_{xz}M_z^2+2\gamma^xk_F^2)^2}- \frac{\gamma_{yy}^x}{(\alpha_y( P )+\beta_{yz}M_z^2+2\gamma^yk_F^2)^2}\right\}\approx 
- \frac{\gamma_{yy}^x \langle \hat k_x^2N_0^\uparrow\rangle}{(\alpha_y( P )+\beta_{yz}M_z^2+2\gamma^yk_F^2)^2}
,\\
g_{2x}^\downarrow\propto\langle \hat k_x^2N_0^\downarrow\rangle
\left\{\frac{\gamma_{xx}^x}{(\alpha_x+\beta_{xz}M_z^2+2\gamma^xk_F^2)^2}- \frac{\gamma_{yy}^x}{(\alpha_y( P_y )+\beta_{yz}M_z^2+2\gamma^yk_F^2)^2}\right\} \approx 
- \frac{\gamma_{yy}^x \langle \hat k_x^2N_0^\downarrow\rangle}{(\alpha_y( P )+\beta_{yz}M_z^2+2\gamma^yk_F^2)^2}, 
\end{eqnarray}
\end{widetext} 
where we have used the smallness of susceptibility along $x$-direction in respect of susceptibility along $y$-axis
 \cite{Hardy2011}.
According to Eqs. (\ref{susc}) and (\ref{Mz})  the denominators in these expressions are decreasing functions of pressure. Thus, the absolute values of the constant of interband pairing increase with uniaxial pressure.

Thus, all the terms in Eq.(\ref{g}) increase with the uniaxial pressure increase, what results in the growing up the temperature of superconducting transition  shown in Fig.1.

\section{Phase transition
 in magnetic field perpendicular to easy magnetization axis}

The reentrant superconducting state in URhGe arises in high magnetic field along b-axis in vicinity of the first order transition from the ferromagnetic to the paramagnetic state. The change of the phase transition type from  the second to the first order has been described phenomenologically in Ref.5.  With aim to establish the uniaxial pressure dependence of the transtion transformation  we  reproduce this derivation.

The equilibrium magnetization projection along the $y$ direction is given by (\ref{My}).
Substituting  this formula to Eq.(\ref{F22})
we obtain
\begin{eqnarray}
F_M=\alpha_{z}(P_y)M_{z}¥^{2}
+\beta_{z}M_{z}^{4}+\delta_zM_z^6-\frac{1}{4}\frac{H_y^2}{\alpha_y(P_y)+\beta_{yz}M_z^2},
\label{F1}
\end{eqnarray}
that gives after expansion of the denominator in the last term, 
\begin{equation}
F_M=-\frac{H_y^2}{4\alpha_y(P_y)}+\tilde\alpha_{z}M_{z}¥^{2}
+\tilde\beta_{z}¥M_{z}¥^{4}+\tilde\delta_zM_z^6+\dots,
\label{F2}
\end{equation}
where
\begin{eqnarray}
&\tilde\alpha_{z}=\alpha_{z0}(T-T_{c0}(P_y))+\frac{\beta_{yz}H_y^2}{4(\alpha_y(P_y))^2},\\
&\tilde\beta_{z}=\beta_z-\frac{\beta_{yz}}{\alpha_y(P_y)}\frac{\beta_{yz}H_y^2}{4(\alpha_y(P_y))^2}\\
&\tilde\delta_{z}=\delta_z+\frac{\beta_{yz}^2}{(\alpha_y(P))^2}\frac{\beta_{yz}H_y^2}{4(\alpha_y(P_y))^2}
\end{eqnarray}

We see that under a magnetic field perpendicular to the direction of spontaneous magnetization  the Curie temperature
 decreases as
\begin{eqnarray}
T_c=T_c(P_y,H_y)=T_{c0}(P_y)-\frac{\beta_{yz}H_y^2}{4\alpha_{z0}(\alpha_y( P_y))^2}\nonumber\\
=T_{c0}-\frac{A_zP_y}{\alpha_{z0}}-\frac{\beta_{yz}H_y^2}{4\alpha_{z0}(\alpha_{y}-|A_y|P_y)^2}.
\label{Cur}
\end{eqnarray}
Thus,
 at a finite field  $H_y$  the Curie temperature suppression by uniaxial pressure occurs much faster than in the field absence.

The coefficient $\tilde\beta_z$ also decreases with $H_y$ and reaches  zero at
\begin{equation}
H_y^{cr}=\frac{2(\alpha_y(P_y))^{3/2}\beta_z^{1/2}}{\beta_{yz}}.
\label{R}
\end{equation}
At this field under fulfillment the  condition,
\begin{equation}
\frac{\alpha_{z0}\beta_{yz}T_{c0}}{\alpha_y(P)\beta_z}>1
\end{equation}
the Curie temperature (\ref{Cur}) is still positive and at $$H_y>H_y^{cr}(P)$$  phase transition from a paramagnetic to a ferromagnetic state becomes the transition of the first order
(see Fig.9 in the Ref.2). The point $(H_y^{cr},T_c(H_y^{cr}))$ on the line  paramagnet-ferromagnet phase transition is a tricritical point.
The pressure dependence  
\begin{equation}
H_y^{cr}(P_y)\propto\left(1-\frac{|A_y|}{\alpha_y}P_y\right)^{3/2}
\end{equation}
roughly corresponds to the observed experimentally \cite{Braithwaite2016} pressure dependence $H_R( P_y )$. An uniaxial pressure enhancement decreases the field of the first order phase transition from ferromagnetic to paramagnetic state.

The minimization of the free energy Eq. (\ref{F2})  gives the value of the order parameter in the ferromagnetic state,
\begin{equation}
M_z^2=\frac{1}{3\tilde\delta_z}[-\tilde\beta_z+\sqrt{\tilde\beta_z^2-3\tilde\alpha_z\tilde\delta_z}].
\end{equation}
The minimization of the free energy in the paramagnetic state,
\begin{equation}
F_{para}=\alpha_y(P_y)M_y^2-H_yM_y
\label{p}
\end{equation}
in respect $M_y$ gives the equilibrium value of magnetization projection on axis $y$ in paramagnetic state,
\begin{equation}
M_y=\frac{H_y}{2\alpha_y(P)}.
\end{equation}
Substitution back in Eq. (\ref{p}) yields the equilibrium value of free energy in the paramagnetic state,
\begin{equation}
F_{para}=-\frac{H_y^2}{4\alpha_y(P_y)}.
\end{equation}
On the line of the phase transition of the first order from the paramagnetic to ferromagnetic state
determined by  the equations \cite{StatPhysI}
\begin{equation}
F_M=F_{para},~~~~~\frac{\partial F_M}{\partial M_z}=0
\end{equation}
the order parameter $M_z$ has the  jump (see Fig.10 in the Ref.2) from
\begin{equation}
 M_z^{\star^2}=-\frac{\tilde\beta_z}{2\tilde\delta_z}.
 \label{jump}
\end{equation}
in the ferromagnetic state to zero in the paramagnetic state.
Its substitution  back in  equation $F_M=F_{para}$ gives the equation of the first-order transition line,
\begin{equation}
4\tilde\alpha_z\tilde\delta_z=\tilde\beta_z^2,
\label{line}
\end{equation}
that is
\begin{equation}
T^\star=T^\star(H_y)=T_{c0}-\frac{\beta_{yz}H_y^2}{4\alpha_{z0}(\alpha_y(P_y))^2}+\frac{\tilde\beta_z^2}{4\alpha_{z0}\tilde\delta_z}.
\end{equation}

The corresponding  jump of $M_y$ (see Fig.10 in Ref.1) is given by 
\begin{equation}
M_y^\star=M_y^{ferro}-M_y^{para}=\frac{H_y}{2(\alpha_y( P_y)+\beta_{yz}M_z^{\star^2})}-\frac{H_y}{2\alpha_y(P_y)}.
\label{jumpy}
\end{equation}

\section{Concluding remarks}

We have shown that both the Curie and the superconducting critical temperature are changed with the pressure applied to URhGe specimen. The functional pressure dependences can be established from general phenomenological considerations. Whereas the direction of changes
must be chosen by comparison with the experimental findings according to which   uniaxial  in b-direction and  hydrostatic pressure act in the opposite sense:   the first  suppresses the ferromagnetism and enhances superconductivity, the second stimulates ferromagnetism and suppresses superconducting state. The phenomenological approach allows to point out two mechanisms of superconducting temperature raising. They originates from stimulation by the uniaxial stress  both intraband and interband amplitudes  of triplet Coper pairing.

The magnetic field in b-direction decreases the Curie temperature and leads to the transformation of the ferro-para phase transition from the second to the first order.
The pairing interaction in vicinity of the first order transition from ferromagnetic to paramagnetic state caused by field $H_y$ is  strongly increased in comparison with its zero field value. This effect proves to be stronger than 
the orbital suppression of superconductivity by magnetic field and leads to the re-appearance of superconducting state at field of the order 10 Tesla. Here we have demonstrated that in presence of an uniaxial pressure along b-axis 
 the process of the ferromagnetism  suppression by magnetic field occurs much faster. The field induced transformation of the second to the first order
 ferro-para phase transition occurs  at much lower field $H_y^{cr}( P_y)$ values.
 The effect is so strong that even small uniaxial stress causes the coalescence 
 \cite{Braithwaite2016} of superconducting and the re-entrantant superconducting area in the $(H_y,T)$ phase diagram shown in Fig.1.

The shape of superconducting region drown in Fig.1b looks similar to the upper critical field temperature dependence 
$H_{c2}^b(T)$ in the other ferromagnetic superconductor UCoGe.  Unlike to URhGe  the hydrostatic pressure applied to UCoGe suppresses ferromagnetism and stimulate superconductivity
\cite{Aoki14}. This allows us to speculate that in the case of UCoGe  an uniaxial pressure applied along b-axis  can  transform
$S$-shape $H_{c2}^b(T)$ curve in  two separate superconducting and reentrant superconducting regions like it is in URhGe in the absence of uniaxial pressure.

\acknowledgments

I am indebted to D.Braithwaite and J.-P. Brison  for the helpful discussions of  their recent experimental results.

\appendix
\section{}

 The explicit expressions for $A_z$ and $A_y$ coefficients through the elastic and magneto-elastic moduli are
  \begin{eqnarray}
A_z=p_x(b_1+B_1)+ p_y(b_2+B_2)+p_z(b_3+B_3),\\A_y=q_x(b_1+B_1)+ q_y(b_2+B_2)+q_z(b_3+B_3),
\end{eqnarray}

\begin{widetext}
 \begin{eqnarray}
B_1=\lambda_xa_1b_1+\lambda_ya_2b_2+\lambda_za_3b_3+\lambda_{xy}(a_1b_2+a_2b_1)+\lambda_{xz}(a_1b_3+a_3b_1)+\lambda_{yz}(a_2b_3+a_3b_2),\nonumber\\
B_2=\lambda_xb_1^2+\lambda_yb_2^2+\lambda_zb_3^2+2\lambda_{xy}b_1b_2+2\lambda_{xz}b_1b_3+2\lambda_{yz}b_2b_3,\\
B_3=\lambda_xb_1c_1+\lambda_yb_2c_2+\lambda_zb_3c_3+\lambda_{xy}(b_1c_2+b_2c_1)+\lambda_{xz}(b_1c_3+b_3c_1)+\lambda_{yz}(b_2c_3+b_3c_2),\nonumber
\end{eqnarray}
and
\begin{eqnarray}
a_1=D^{-1}(\lambda_{yz}^2-\lambda_y\lambda_z),~~~~a_2=D^{-1}(\lambda_{xy}\lambda_z-\lambda_{xz}\lambda_{yz}),~~~~a_3=D^{-1}(\lambda_{y}\lambda_{xz}-\lambda_{xy}\lambda_{yz}),\nonumber\\
b_1=D^{-1}(\lambda_{xy}\lambda_z-\lambda_{yz}\lambda_{xz}),~~~~b_2=D^{-1}(\lambda_{xz}^2-\lambda_x\lambda_z),~~~~b_3=D^{-1}(\lambda_x\lambda_{yz}-\lambda_{xz}\lambda_{xy}),\\
c_1=D^{-1}(\lambda_{y}\lambda_{xz}-\lambda_{xy}\lambda_{yz}),~~~~c_2=D^{-1}(\lambda_{x}\lambda_{yz}-\lambda_{xy}\lambda_{xz}),~~~~c_3=D^{-1}(\lambda_{xy}^2-\lambda_x\lambda_y),\nonumber\\
D=\lambda_x\lambda_y\lambda_z+2\lambda_{xz}\lambda_{xy}\lambda_{yz
}-\lambda_{xz}^2\lambda_{y}-\lambda_{yz}^2\lambda_{x}-\lambda_{xy}^2\lambda_z.~~~~~~~~~~~~~~
\end{eqnarray}
\end{widetext}

\begin{figure}[p]
\includegraphics
[height=.8\textheight]
{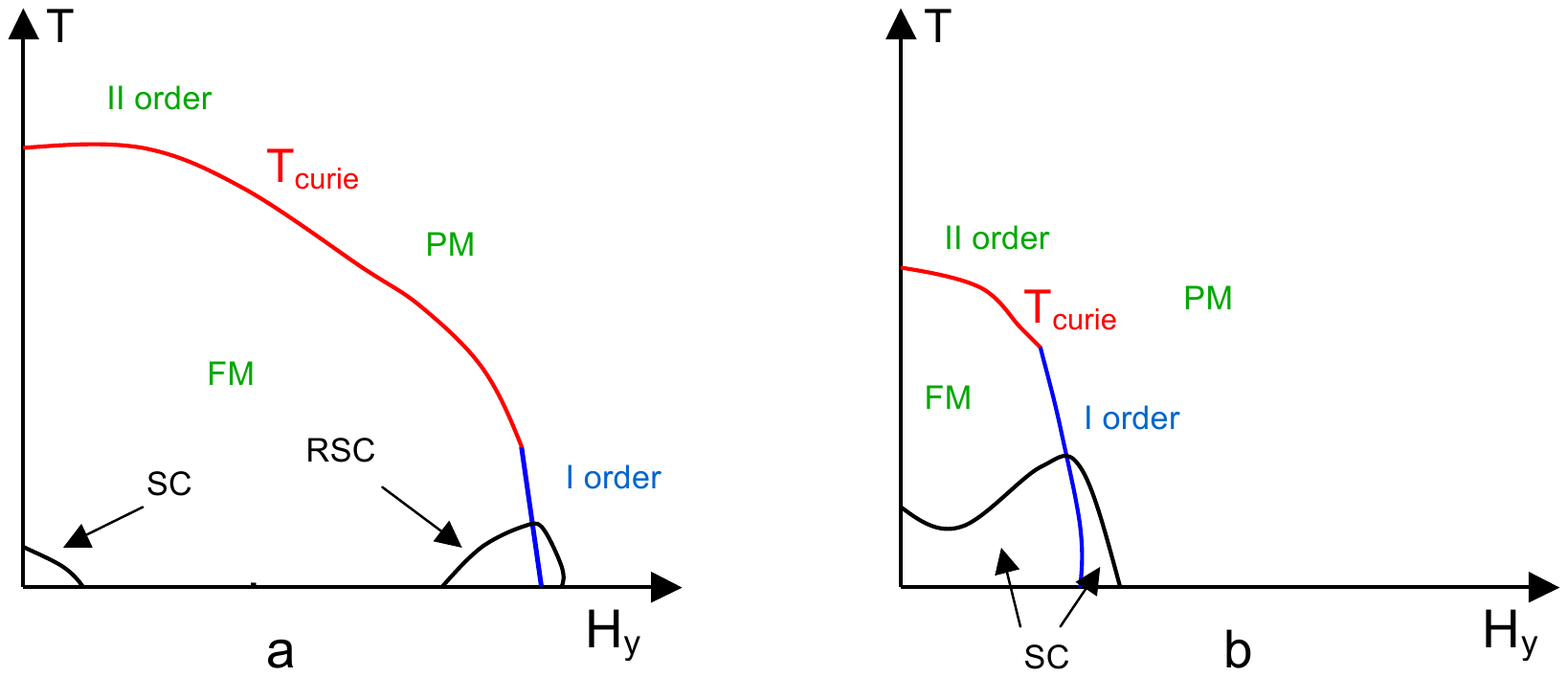}
 \caption{(Color online) 
 Schematic phase diagrams of  URhGe in a magnetic field along b-axis perpendicular to the spontaneous magnetization direction: $(a)$ at ambient pressure and  $(b)$ at strong enough uniaxial stress along b-direction.  PM and FM denote paramagnetic and ferromagnetic states. 
 SC and RSC are the superconducting and reentrant superconducting states correspondingly.
 The red line is the Curie temperature. The blue line is the line of the first order transition.
  }
\end{figure}


\begin{thebibliography}{220}

\bibitem{Aoki14} D. Aoki and J. Flouquet, J. Phys. Soc. Jpn. {\bf 83}, 061011 (2014).

\bibitem{Mineev2016}V. P.Mineev, Usp. Fiz. Nauk {\bf 187}, 129 (2017) [Phys.-Usp. {\bf 60}, 121
(2017)], arXiv:1605.07319.

\bibitem{Levy2005}  Levy F,  Sheikin I,  Grenier B, Huxley A D  {\it Science} {\bf 309} 1343 (2005)

\bibitem{Hardy2005}F. Hardy, A.D. Huxley, Phys. Rev. Lett. {\bf 94}, 247006 (2005).

\bibitem{Mineev2015} V.P.Mineev,  Phys. Rev. B {\bf 91} 014506 (2015).

\bibitem{Hardy}F. Hardy, A. Huxley, J. Flouquet, B. Salce, G. Knebel, D. Braithwaite,
D. Aoki, M. Uhlarz, C. Pfleiderer, Physica B {359-361}, 1111 (2005).

\bibitem{Miyake2009} A.Miyake, D.Aoki, and J. Flouquet, J.Phys. Soc. Jpn. {\bf 78}, 063703 (2009).

\bibitem{Braithwaite2016} D.Braithwaite, D.Aoki, J.-P.Brison, J.Flouquet, G.Knebel, A.Pourret,
{\it to be published} (2017).

\bibitem{TheorEl}L.D. Landau and E.M. Lifshitz, {\it  Theory of Elasticity,  Course of Theoretical Physics Vol.VII.} Oxford: Butterworth-Heinemann (1986).

\bibitem {Mineev2011} V.P.Mineev, Phys. Rev. B {\bf 83} 064515 (2011).

\bibitem{Hardy2011} F.Hardy, D.Aoki, C.Meingast, P.Schweiss, P.Burger, H. v.Loehneysen, and J.Flouquet,
Phys.Rev.b {\bf 83}, 195107 (2011).


\bibitem{StatPhysI} L.D.Landau and E.M.Lifshitz, 
{\it Statistical Physics, Course of Theoretical Physics Vol V.} Oxford:  Butterworth-Heinemann,1995)



\end{thebibliography}
\end{document}